\documentclass[journal]{IEEEtran}
\ifCLASSINFOpdf
\else
\fi

\usepackage{graphicx}
\usepackage{booktabs}
\usepackage{hyperref}
\usepackage[ruled]{algorithm2e}
\usepackage{algorithmic}

\usepackage{amsmath}
\usepackage{amssymb}
\usepackage{mathtools}
\usepackage{amsthm}
\usepackage{amsfonts}
\usepackage{textcomp}
\usepackage[T1]{fontenc}
\usepackage{bm}       
\usepackage{multirow}
\usepackage{array}
\usepackage{marvosym}
\usepackage{ragged2e}
\usepackage{enumerate}
\usepackage{enumitem}
\usepackage{tikz,lipsum,lmodern}
\usepackage[most]{tcolorbox} 
\usepackage{scalerel}
\usepackage[capitalize,noabbrev]{cleveref}
\theoremstyle{plain}
\newtheorem{theorem}{Theorem}[section]
\newtheorem{proposition}[theorem]{Proposition}

\theoremstyle{definition}

\theoremstyle{remark}

\newcommand{\Hyb}[1]{\quad \bm{\mathrm{Hyb}}_{#1}}

\newcommand{\scheme}{FedCAPrivacy\xspace}

\begin{document}
%
\title{\scheme: Privacy-Preserving Heterogeneous \\Federated Learning with Anonymous Adaptive Clustering}
%
%
%

\author{Yunan~Wei,
        Shengnan~Zhao,
        Chuan~Zhao$^*$,
        Zhe~Liu,
        Zhenxiang~Chen, 
        and~Minghao~Zhao
\thanks{Yunan Wei is with the Shandong Key Laboratory of Ubiquitous Intelligent Computing, University of Jinan, Jinan 250022, China and Quan Cheng Laboratory, Jinan 250103, China (email: wyn$@$stu.ujn.edu.cn).}
\thanks{Shengnan Zhao is with Quan Cheng Laboratory, Jinan 250103, China (email: zsn.sdu$@$gmail.com).}
\thanks{Chuan Zhao is with Quan Cheng Laboratory, Jinan 250103, China (email: ise\_zhaoc$@$ujn.edu.cn).}
\thanks{Zhe Liu is with Zhejiang Laboratory, Hangzhou 311121, China (email: zhe.liu$@$zhejianglab.com).}
\thanks{Zhenxiang Chen is with the Shandong Key Laboratory of Ubiquitous Intelligent Computing, University of Jinan, Jinan 250022, China (email: zxchen$@$ujn.edu.cn).}
\thanks{Minghao Zhao is with the School of Data Science and Engineering, East China Normal University, Shanghai 200050, China (mhzhao$@$dase.ecnu.edu.cn).}
\thanks{$*$ Chuan Zhao is the corresponding author. }
}

%
%

\markboth{Journal of \LaTeX\ Class Files,~Vol.~XX, No.~XX}%
{Shell \MakeLowercase{\textit{et al.}}: Bare Demo of IEEEtran.cls for IEEE Journals}
%



\maketitle
\begin{abstract}
Federated learning (FL) is a distributed machine learning paradigm enabling multiple clients to train a model collaboratively without exposing their local data.
Among FL schemes, \textit{clustering} is an effective technique addressing the heterogeneity issue (i.e., differences in data distribution and computational ability affect training performance and effectiveness) via grouping participants with similar computational resources or data distribution into clusters.
However, intra-cluster data exchange poses privacy risks, while cluster selection and adaptation introduce challenges that may affect overall performance.
To address these challenges, this paper introduces anonymous adaptive clustering, a novel approach that simultaneously enhances privacy protection and boosts training efficiency.
Specifically, an oblivious shuffle-based anonymization method is designed to safeguard user identities and prevent the aggregation server from inferring similarities through clustering.
Additionally, to improve performance, we introduce an iteration-based adaptive frequency decay strategy, which leverages variability in clustering probabilities to optimize training dynamics.
With these techniques, we build the \scheme; experiments show that \scheme achieves $\sim7\times$ improvement in terms of performance while maintaining high privacy.
\end{abstract}

%
\IEEEpeerreviewmaketitle

\section{Introduction}
\label{sec:intro}

Federated Learning (FL)~\cite{FedAvg} is an emerging machine learning paradigm that enables collaborative model training across multiple decentralized devices without requiring the direct exchange of raw data. 
Recently, FL has been widely adopted in various industries, ranging from medical image analysis~\cite{li2019privacy, liu2021feddg} to autonomous driving~\cite{niknam2020federated}, whenever data privacy and security are paramount concerns.

In federated learning (FL) schemes, {\em heterogeneous federated learning} accounts for and addresses the heterogeneity in data distributions, model architectures, network conditions, and hardware capabilities among participating clients.
 Achieving {\em security, privacy, and desirable performance} simultaneously in heterogeneous federated learning remains challenging, as such heterogeneity imposes additional computational and privacy protection burdens~\cite{ye2023heterogeneous}.

For example, \emph{Device heterogeneity}, normally derived from variations in client hardware capabilities, leads to disparities in training speeds and computational resources. This issue is exacerbated in large-scale FL deployments, where differences in stochastic gradient descent (SGD) iterations among devices introduce inconsistencies in model updates~\cite{huang2023generalizable}.
Variations in data distribution (i.e., non-identically distributed (non-IID) and imbalanced nature of client datasets) can significantly degrade model convergence and generalization performance~\cite{ye2023heterogeneous}. 
Beyond that, for privacy aspects, even DF conceals users' data, recent research~\cite {DLG, wang2024more} finds that model parameters shared by participants can be exploited to infer raw training data, thereby compromising user privacy.
For security aspects, adversarial attacks or transmission errors may distort model updates, leading to compromised training outcomes~\cite{shen2023verifiable, eltaras2023efficient}.

To safeguard participants' privacy, homomorphic encryption (HE) is widely employed as an underlying technology to ensure the confidentiality of model parameters. HE enables the correct aggregation of model parameters without revealing their plaintext. Prior studies~\cite{shen2023verifiable, VerifyNet, liu2021privacy} have also shown the superior performance of HE to achieve secure FL. Besides, cluster techniques~\cite{mishra2024resource,li2022fedhisyn} are widely adopted to deal with heterogeneous problems. Specifically, cluster-based FL groups clients based on device characteristics or data distribution similarities, allowing clients within the same cluster to share common training strategies and enhance model performance.
Despite extensive investigations into the topic of privacy-preserving FL with clustering, existing schemes have ignored the following problems:

1) \emph{Risks of privacy leakage}: The clustering process naturally reveals client similarities that a centralized service provider might exploit to deduce further sensitive information about users.

2) \emph{Communication constraint}: The scalability and availability of systems are greatly affected by inter-node communication, particularly in large-scale FL. The complexity of data exchange between client pairs increases, which likely accounts for the design of most FL schemes under the principle of "no inter-node communication."

3) \emph{Scalability and computational complexity}: In dynamic environments, clustering necessitates continual adjustments as client participation fluctuates, leading to added computational overhead. Furthermore, clustering in large-scale FL systems poses difficulties due to the high dimensionality of client features and the expenses involved in maintaining cluster consistency.

To ensure privacy during cluster processing, we propose an oblivious shuffle-based anonymization technique to prevent the aggregation server from identifying participants. Additionally, we demonstrate that secure clustering in each federated training round offers only marginal performance improvements post-convergence while introducing unnecessary computational overhead. Consequently, we present an adaptive approach to adjust clustering frequency based on iterative decay. Experimental results highlight the robustness of our algorithm across various heterogeneous resources.



In this work, we propose a novel FL framework, called \scheme. In a nutshell, our contributions are summarized as follows:
\begin{itemize}
    \item To the best of our knowledge, we propose the first adaptive frequency with iteration-based decay on clustering that can reduce computational redundancy and improve training efficiency.
    \item We propose an oblivious shuffle-based method for anonymizing user identity to protect participants' identity in case an aggregator infers pairwise similarity during clustering.
    \item We provide a rigorous theoretical analysis of our proposed methods. The experiments conducted on real-world datasets show that our \scheme achieves better model performance while speeding up training in the presence of heterogeneity.
\end{itemize}

\textbf{Paper Organization}: Section~\ref{sec:rw} provides an overview of related works. Section~\ref{sec:pre} outlines the preliminary information and problem statement. We detail our proposed framework in Section~\ref{sec:fw} and analyze the security properties of Fed-PAC in Section~\ref{sec:ana}. Section~\ref{sec:eva} evaluates the performance of our framework. In the end, Section~\ref{sec:clu} concludes our work.
\section{Related Works}
\label{sec:rw}
In recent years, there has been a notable increase in studies addressing privacy protection and security issues in FL~\cite{FedAvg, DLG, zhang2021leakage, geiping2020inverting}. Current research has leveraged advanced techniques such as differential privacy~\cite{luo2024privacy}, homomorphic encryption~\cite{VerifyNet, liu2021privacy}, and secure multi-party computation~\cite{bonawitz2017practical, shen2023verifiable} to ensure data confidentiality.

Several approaches have been proposed to address the inherent heterogeneity present within the field of federated learning (FL). These include the utilization of resource selection methods~\cite{FedMCCS}, the mitigation of model convergence bias~\cite{FedProx, SCAFFOLD}, and the implementation of federated distillation methods~\cite{hu2020gfl, huang2022learn}. These approaches aim to mitigate the adverse effects of non-IID data distributions and resource disparities between clients.

In the context of clustering methods, various studies have applied clustering techniques~\cite{lee2024softcluster, li2022fedhisyn, luo2024privacy, ghosh2020efficient} to improve training efficiency and overall effectiveness in FL. However, many of these methods ~\cite{liu2021privacy, li2022fedhisyn} neglect to prioritize the safeguarding of participant similarity, a factor that may result in privacy risks. 
\section{Preliminaries}
\label{sec:pre}

\subsection{Federated Learning}
\label{subsec:fl}
FL usually consists of a central server $S$ and $n$ devices. At the beginning of each iteration, a subset of devices denoted as $P$, is selected to participate in the training. The training process is described as follows:

1) Initialization: The central server initializes and broadcasts the global model parameters $\omega_g$.

2) Local Training: Each participating device $p_i \in P$ receives $\omega_g^r$ at round $r$ and updates the local model parameters $\omega_i$ by performing $E$ epochs of stochastic gradient descent (SGD) on its local dataset $D_i$:
\begin{equation}
    \omega_i^{r+1} = \omega_i^r - \eta\nabla\ell(\omega_i^r;D_i)
\end{equation}
where $\eta$ is the learning rate and $\ell(\cdot)$ is the loss function.

3) Model Aggregation: For each $p_i \in P$, the updated model parameters $\omega_i^{r+1}$ are uploaded to $S$. All the user model parameters are aggregated in $C$ to update the global model:
\begin{equation}
    \omega_g^{r+1} = \sum_{i=1}^{|P|} \frac{d_i}{d} \omega_i^{r+1}
\end{equation}
where $||P||$ is the number of participating devices, $d_i$ represents the size of the local dataset of $i^{th}$ device and $d$ is the number of the total training samples across all devices in this round. Repeat steps 2 and 3 for $E$ rounds or until the global model converges. 

\subsection{Homomorphic Encryption}
HE allows computation directly on encrypted data without requiring access to a secret key. In this paper, we implement FL that improves privacy by exploiting the Paillier cryptosystem~\cite{paillier}. In general, the cryptosystem consists of the following three algorithms:
\begin{itemize}
    \item KeyGen$(1^\kappa) \to (pk, sk)$: Input a security parameter \( \kappa \), and choose two large prime numbers \( p \) and \( q \) each to calculate \( N = pq \) and \( \mathcal{L} = \textit{lcm}(p-1, q-1) \). And then select \( g \in \mathbb{Z}_{N^2}^* \) such that \( g = N + 1 \), compute \( \mu = (\textit{L}(g^\mathcal{L} \mod n^2))^{-1} \mod N \), where \( \textit{L}(x) = \frac{x-1}{n} \). The public key $pk$ is \( (N, g) \) and the private key $sk$ is \( (\mathcal{L}, \mu) \).
    \item Enc$(pk, m) \to c$:  To encrypt a message \( m \in \mathbb{Z}_N \), select a random \( r \in \mathbb{Z}_n^* \) and compute the ciphertext \( c = g^m \cdot r^N \mod N^2 \).
    \item Dec$(sk, c) \to m$: To decrypt a ciphertext \( c \in \mathbb{Z}_{N^2}^* \), compute \( m = \textit{L}(c^\mathcal{L} \mod N^2) \cdot \mu \mod N \).
\end{itemize}
Given two plaintexts $c1$ and $c2$, a constant $r$, and a public key $pk$, we have:

1) \emph{Addition of plaintexts:} Given \( c_1 = Enc(pk, m_1) \) and \( c_2 = Enc(pk, m_2) \), the product \( c_1 \cdot c_2 \mod n^2 \) decrypts to \( m_1 + m_2 \mod n \).

2) \emph{Scalar multiplication:} Given \( c = Enc(m, r) \) and a scalar \( k \in \mathbb{Z}_n \), \( c^k \mod n^2 \) decrypts to \( k \cdot m \mod n \).

\section{\scheme Framework}
\label{sec:fw}



\subsection{Technical Overview}

\begin{figure*}[!ht]
\vskip 0.2in
\begin{center}
\centerline{\includegraphics[width=2\columnwidth]{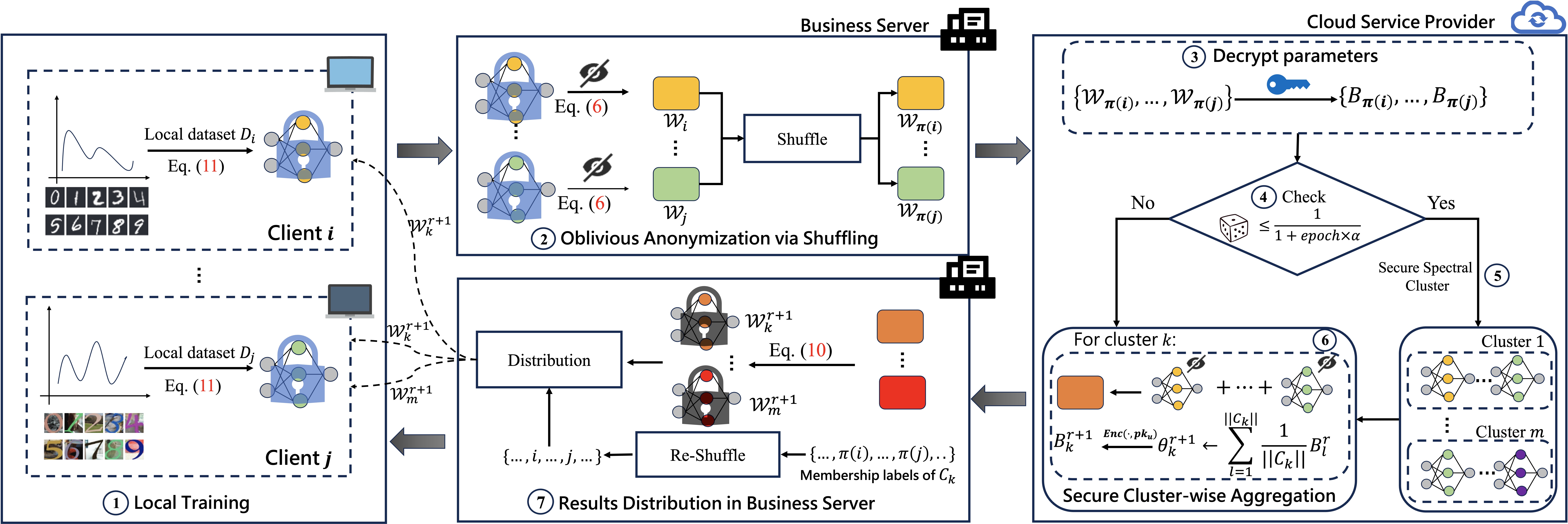}}
\caption{Illustration of \scheme workflow.}
\label{fig: workflow}
\end{center}
\vskip -0.2in
\end{figure*}

\scheme consists of three main components: (1) a business server (BS) to collect and aggregate model parameters; (2) a cloud service provider (CSP) that clusters model parameters based on similarity; and (3) devices with heterogeneous data (as shown in ~\cref{fig: workflow}). 


\emph{Oblivious Identity Anonymization via Shuffling.}
The aggregator may infer associations between participants’ identities and the ciphertext indices they upload, thereby revealing links between data and senders and exposing similarities among participants during clustering. Without adequate safeguards, this could lead to unintentional privacy breaches by the aggregation server. Unlike prior work that overlooks the privacy risks posed by participant similarities, we stress the importance of protecting participant identities during clustering to prevent the CSP from deducing such relationships. To mitigate this risk, we propose anonymizing participant identities by shuffling the ciphertext set, effectively masking similarities among participants and enhancing privacy throughout the clustering process.

We assume that the ciphertexts set the BS received is $\mathbf{\mathcal{W}} = \{[\mathcal{W}_1]_{pk_s}, ...,[\mathcal{W}_i]_{pk_s} \}$. Then the oblivious identity anonymization proceeds as follows. First, the BS selects a random number $\mu$ to blind the set $\mathbf{\mathcal{W}}$ to $\mathbf{\mathcal{B}} = \{[\mathcal{B}_1]_{pk_s}, ...,[\mathcal{B}_i]_{pk_s} \}$. Then the BS permutes the set $\mathbf{\mathcal{B}}$ to $\pi(\mathbf{\mathcal{B}}) = \{[\mathcal{B}_{\pi(1)}]_{pk_s}, ...,[\mathcal{B}_{\pi(i)}]_{pk_s}\}$ by a randomly chosen permutation $\pi$. With such a design, \scheme can efficiently protect participants' identities without introducing the amount of computational overhead. The overall method is shown in Lines 14-18 in~\cref{alg: workflow}.


\begin{figure}[t]
\vskip 0.2in
\begin{center}
\centerline{\includegraphics[width=\columnwidth]{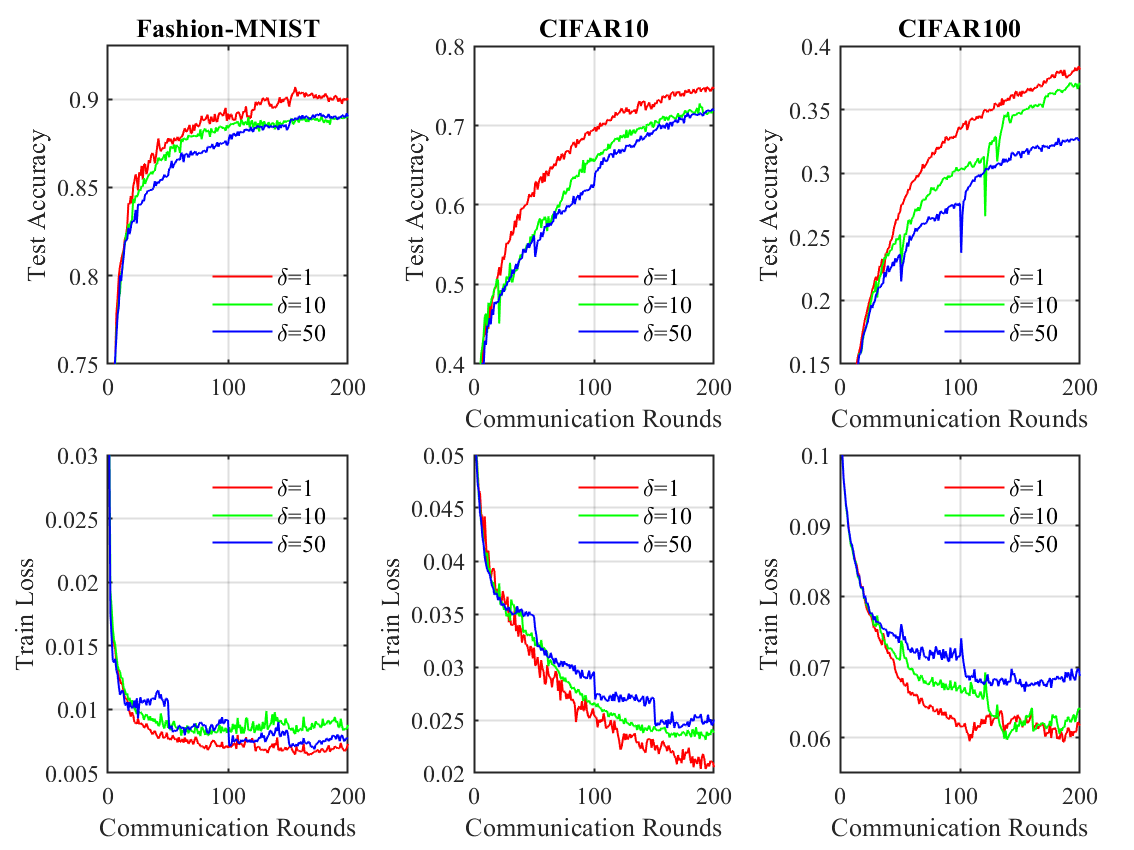}}
\caption{Influence of the frequency of clusters for heterogeneous resources.}
\label{fig:cls-freq}
\end{center}
\vskip -0.2in
\end{figure}

\emph{Dynamic frequency with iteration-based decay on clustering (DFD).} 
Reducing clustering frequency intuitively decreases computational overhead. We evaluate scenarios where clustering is executed every 1, 10, and 50 rounds to investigate its effect on model performance. As shown in ~\cref{fig:cls-freq}, lower clustering frequencies lead to reduced model accuracy and increased loss values. During the convergence phase, accuracy exhibits significant fluctuations, while the loss function decreases progressively with each clustering step. These effects become negligible after the model converges. Excessive clustering introduces computational redundancy, particularly after convergence, while reducing clustering frequency effectively lowers costs and accelerates training. However, overly sparse clustering may result in local models primarily learning from similar data distributions, leading to suboptimal solutions. This also accounts for the faster initial convergence observed with a clustering interval of 50. Nevertheless, as clustering frequency decreases, disparities among cluster models widen, hindering the formation of new clusters and limiting the model’s ability to generalize effectively, ultimately constraining final accuracy.

According to our investigation of the effect of clustering frequency on model performance and training efficiency, we introduce an adaptive adjustment for clustering frequency to reduce computational redundancy. Specifically, in the $r$ round, the probability of clustering is determined by $Pr = \frac{1}{1+\alpha\cdot r}$, where $\alpha$ is the decay factor that controls how fast the probability of clustering decay is as the number of iterations grows. 

\subsection{Construction of \scheme}

\subsubsection{Setup Phase}

A key generate center (KGC) is required to create a pair of key $(pk_s, sk_s)$ of the Paillier cryptosystem for the CSP and $(pk_u, sk_u)$ for all participants, where $sk_s$ is only kept by the CSP and $sk_u$ is stored locally in each device. 
The BS provides the initial global weight ${\omega}_g^0$ and assigns the index as the distinct identifier of the participants $PID$. Every $R$ time, the server aggregates the updated models it receives from all devices to complete a training round.

\begin{algorithm}[!ht]
    \LinesNumbered
    \caption{The training process of \scheme} 
    \label{alg: workflow}
    \begin{algorithmic}
        \STATE {\bfseries Input:}
        Number of model clusters $c$, set of device $M$, the time interval at which the server receives parameters $R$, time of device i to complete the local training $t_i$, encryption key pair $(pk_u, sk_u)$, a shuffle function $\pi(\cdot)$, an Initialized global model parameters $\omega_g^0$, parameter of Gaussian-RBF $\gamma$, encryption key pair $(pk_s, sk_s)$
        \SetKwFunction{FShuffle}{Shuffle}
        \SetKwFunction{FSSC}{SecureCluster}
        \SetKwProg{Fn}{Function}{:}{end} 
        \STATE $\omega_{0,1,...,n-1} \gets \omega_g^0$
        \FOR{iteration $r = 1, 2, ...$}
            \STATE Randomly select a subset $P$ from device set $M$
            \STATE $R_{p_1,p_2,...,p_n}\gets R$
            \REPEAT
                \FOR{$p_i \in P$ in parallel}
                    \STATE $R_{p_i} = R_{p_i} - t_i$
                    \STATE $\omega^{r+1}_i=\omega^r_{i,g}-\eta\triangledown \ell(\omega)$
                    \STATE $[\mathcal{W}_i]_{pk_s} \gets $ \textit{Enc($\omega_i, pk_s$)}
                \ENDFOR
            \UNTIL{$R_{p_i} \leq 0$}
            \STATE Randomly select a blinding factor $\mu$
            \STATE $\textbf{$\mathcal{B}$}=\textbf{$\mathcal{W}$} \cdot [\mu]_{pk_s}$
            \STATE $\pi(\textbf{$\mathcal{B}$})\gets
            \FShuffle(\textbf{$\mathcal{B}$})$
            \STATE $\pi(B) \gets$ \textit{Dec}($\pi(\mathcal{B}), sk_s$)
            \IF {a random value $(\in [0, 1)) \leq 1/(1 + \alpha \cdot r)$ or $r==1$}   
                \STATE $\varphi \gets \FSSC(\pi(B), c, \gamma)$
            \ENDIF
            \FOR {$k = 0,1,...,c-1$}
                \STATE Aggregate model parameters within cluster $C_k$ : $B_k = \frac{1}{\|C_k\|} \sum_{l=1}^{||C_k||} B_l$ \;
                \STATE $[B_k]_{pk_u} \gets $ \textit{Enc}($B_k, pk_u$)
            \ENDFOR

            \FOR {each $[B_k]_{pk_u}$}
                \STATE $[\mathcal{W}_k]_{pk_u} = [B_k]_{pk_u} \cdot [\frac{1}{||C_k||}\cdot[\mu]_{pk_u}]^{-1}$
            \ENDFOR
            \STATE Distribute $[\mathcal{W}_k]_{pk_u}$ to $p_{\pi^{-1}(j)}$ where $\varphi_{k,j}=1$
  
            \FOR{$p_i \in P$ in parallel}
                \STATE $\omega_{i,g}= \frac{1}{||C_{p_i}||}\sum^{||C_{p_i}||}_{l=1} \omega_l$
            \ENDFOR
        \ENDFOR
        \STATE \Fn{\FShuffle{$\mathcal{B}$}}{
            Swap $\mathcal{B}_i \leftrightarrow \mathcal{B}_{\pi(i)}$\;
        }
        \STATE \Fn{\FSSC{$B, c, \gamma$}}{
            Call~\cref{alg: PSC}\;
        }
    \end{algorithmic}
\end{algorithm}

\subsubsection{Training at Local devices}

\textit{Step \uppercase\expandafter{\romannumeral1}}: Participant $p_i \in P$ selects a subset of the local dataset and applies an update as follows:  
\begin{equation}
    \omega_i^{r+1}=\omega_i^r-\eta\triangledown \ell_i(\omega)
\end{equation}
where $\ell(\cdot)$ is the local loss function and $r$ denotes the round number. 
Due to the essential requirements of the cryptographic primitives, each entry $\omega_{i,x}$ should be encoded into integer form as follows: 
\begin{equation}
    \omega_i \gets \{\lceil2^\rho \cdot \omega_{i,x} \rfloor\}^{x=m}_{x=1}
\end{equation}
where $\lceil \mathbb{X} \rfloor$ represents the nearest integer to the real number $\mathbb{X}$, $\rho$ is the precision. 
$p_i$ encrypts $\omega_i$ with CSP's public key $pk_s$ as :
\begin{equation}
     [\mathcal{W}_i]_{pk_s}= \textit{Enc}(\omega_i, pk_s)
\end{equation}
where $[\cdot]_{pk}$ denotes the ciphertext encrypted with the public key $pk$. 
$p_i$ generates a data tuple
$T_i=([\mathcal{W}_i]_{pk_s},PID,TS)$, where $PID=i,0\leq i < n$, $TS$ is a timestamp. 
Then $p_i$ calculates the signature $\zeta_i = H(T_i)^{\lambda_i}$ and sends $(T_i, \zeta_i)$ to BS.

\begin{algorithm} [t]
 \caption{Oblivious Spectral Clustering} \label{alg: PSC}
 \begin{algorithmic}
  \LinesNumbered
     \STATE {\bfseries Input:}Set of Model parameters $\{{B}\}$, numbers of spectral clustering $c$, parameter of Gaussian-RBF $\gamma$
     \STATE {\bfseries Output:}Clustering label matrix $\varphi$
     \FOR{$i = 1,2,..., n$}
        \FOR{$j =1,2,...,n$}
             \STATE $s_{i,j} = \exp(-\frac{\|B_i - B_j\|_2}{2\gamma})$\;
        \ENDFOR
     \ENDFOR
     \STATE $A = [s_{i,j}]_{n\times n}$
     \STATE $D \gets diag(A)$
     \STATE $L = D - A$
     \STATE $L_{norm} = D^{-1/2} L D^{-1/2}$
     \STATE $U \gets eigenvectors(L_{norm})[:, -c:]$
     \FOR{$i = 1$ \KwTo $n$}
        \STATE$U[i,:] = \frac{U[i,:]}{\|U[i,:]\|}$
     \ENDFOR
     \STATE $C \gets cluster(U, c)$
     \STATE $E \gets eigenvectors(A)[:, -c:]$
     \STATE $E_{norm} \gets normalize(E)$
     \FOR{$k = 0$ \KwTo $c-1$}
        \STATE $\varepsilon_k \gets mean(E_{norm}[labels == k])$ where $labels \in C$
     \ENDFOR
     \FOR{$j=0$ \KwTo $n-1$}
        \FOR{$k=0$ \KwTo $c-1$}
            \STATE $a_{jk} = \exp(-\frac{\|E_{norm}[j] - \varepsilon_k\|_2}{2\gamma})$\;
        \ENDFOR
        \STATE $\epsilon_j^r \gets mean(a_j)$
        \STATE $\varphi_{jk} \to 1$ if $a_{jk} \geq \epsilon_j^r$ else $0$
     \ENDFOR
 \end{algorithmic}
\end{algorithm}

\subsubsection{Secure Cluster-wise Aggregation}
    
\textit{Step \uppercase\expandafter{\romannumeral2}}: Upon receiving the set of ciphertexts $\{[\mathcal{W}_i]_{pk_s}\}^{i=n-1}_{i=0}$ 
the BS randomly selects a blinding factor $\mu$ to obscure $[\mathcal{W}_i]_{pk_s}$ as follows:
\begin{equation}
    [\mathcal{B}_i]_{pk_s} = [\mathcal{W}_i]_{pk_s} \cdot [\mu]_{pk_s}
\end{equation} 
where $i\in [0,n)$. Then the BS shuffles the $\{\mathcal{B}_i\}^{i=n}_{i=1}$ to $\{\mathcal{B}_j\}^{j=n}_{j=1}$, where $j=\pi(i)$. 
Ultimately, BS sends the perturbed set $\{\mathcal{B}_j\}^{j=n}_{j=1}$ to the CSP.

\textit{Step \uppercase\expandafter{\romannumeral3}}: 
The CSP first decrypts the ciphertexts of model parameters to obtain the blinded model parameters as follows:
\begin{equation}
    B_j = Dec([\mathcal{B}_j]_{pk_s}, sk_s)
\end{equation}

\textit{Step \uppercase\expandafter{\romannumeral4}}: The CSP selects a random to determine whether to conduct clustering at the current round.
If so, the similarity matrix $S$ of the parameters (lines 1 in~\cref{alg: PSC}) is calculated, where the similarity is quantified by the Gaussian radial basis function (RBF) for pairwise perturbed model parameters:
\begin{equation}
     s_{l,j} = \exp\left(-\frac{\|B_l - B_j\|_2}{2\gamma}\right)
\end{equation}
where $\gamma$ is the scale parameter determining the width of the RBF. 
CSP obtains cluster labels for model parameters (Lines 2-8 in Alg.~\ref{alg: PSC}) and utilizes the spectral embedding method to construct the clustering label matrix $\varphi$ (Lines 9-17 in Alg.~\ref{alg: PSC}), where \(a_{jk}\) indicates the affinity of \(\mathcal{B}_j\) to the cluster center \(\varepsilon_k\) of cluster \(C_k\), with \(1 \leq j \leq n\) and \(1 \leq k \leq c\). For each blinded parameter \(B_j\), the affinity budget \(\epsilon_j\) is determined by calculating the average affinity of \(B_j\) across all clusters. If \(a_{jk} \geq \epsilon_j\), then \(B_j\) is allocated to cluster \(C_k\), setting \(\varphi_{jk}=1\) (Lines 13-17 in Alg.~\ref{alg: PSC}). 

\textit{Step \uppercase\expandafter{\romannumeral5}}:
Then CSP aggregates model parameters within each cluster as follows(Line  in Alg.~\ref{alg: workflow}):   
\begin{equation}
    B_{C_k}^{r+1} = \frac{1}{\|C_k\|} \sum_{j=1}^{\|C_k\|} B_j^r
\end{equation}
where $B_{C_k}^{r+1}$ is the cluster-wise aggregated model of $C_k$ with obscurity, and $\|C_k\|$ is the number of participating devices in cluster $C_k$.
Finally, the CSP transmits the clustering label matrix \(\varphi\) and the set of the ciphertexts of cluster-wise aggregated models $\{[B_{C_k}]_{pk_u}\}_{k=1}^{k=c}$ to the BS.

\textit{Step \uppercase\expandafter{\romannumeral6}}: The BS first eliminates the obscurity of $[B_{C_k}]_{pk_u}$ as follows:
\begin{equation}
    [\mathcal{W}_{C_k}]_{pk_u} = [B_{C_k}]_{pk_u} \cdot [\frac{1}{||C_k||} \cdot [\mu_r]_{pk_u}]^{-1}
\end{equation}

Then, BS reveals the participants corresponding to the model parameters in each cluster.
For cluster $C_k$, the real $PID$ of the member $p_j$ equals to $i = \pi^{-1}(j)$. And it distributes $[W_{C_k}]_{pk_u}$ and corresponding signatures to $p_i$.

\textit{Step \uppercase\expandafter{\romannumeral7}}: 
The participant $p_i$ first verifies the model parameters received in clusters. Then, it decrypts the valid parameters to obtain the plaintext parameters {$\omega_{C_k}$} and applies an update as follows:
\begin{equation}
    \omega^{r+1}_{i,g}= \frac{1}{{||C_{p_i}||}}\sum_{k=1}^{||C_{p_i}||} \omega_{C_k}^{r+1}
\end{equation}
where $||C_{p_i}||$ denotes the size of the clusters which $p_i$ belongs to.

\section{Security Analysis}
\label{sec:ana}

We consider servers $S=\{BS, CSP\}$ to interact with a subset of users $U$, and the underlying cryptographic primitives are instantiated with the security parameter $\kappa$. Let \textbf{$REAL_S^{U,\kappa}$} be a random variable that indicates the view of the adversary in the real procession.
\begin{proposition} 
    (Honest but curious security, with curious servers): Given a security parameter $\kappa$, any subset of users $U$ and $S=\{BS, CSP\}$, there is a probabilistic polynomial time (PPT) simulator \textbf{$SIM$} whose output is computationally indistinguishable from \textbf{$REAL_S^{U,\kappa}$}:
    \begin{equation}
    \nonumber
\textbf{$REAL_S^{U,\kappa}$}\approx_c\textbf{$SIM_S^{U,\kappa}$}
    \end{equation}
    where "$\approx_c$" represents computationally indistinguishable.
\end{proposition}
\begin{proof}
    The semantic security of the Paillier cryptosystem is proven in~\cite{paillier}.
    Based on the definition of \textbf{$REAL_S^{U,\kappa}$}, it encompasses all internal states and data received by the entities in $S$ during the execution of the protocol. To prove this proposition, we utilize the standard hybrid argument method as employed in~\cite{liu2021privacy} and~\cite{eltaras2023efficient}. Specifically, for the given security parameter $\kappa$, we define a PPT simulator \textbf{$SIM$} through a series of polynomially many modifications to the random variables in \textbf{$REAL_S^{U,\kappa}$}, showing that the output of \textbf{$SIM$} is computationally indistinguishable from \textbf{$REAL_S^{U,\kappa}$}. 
    Detailed proof is provided below.

    $\Hyb{0}$: We initialize a random variable whose distribution is indistinguishable from \textbf{$REAL_S^{U,\kappa}$}, the joint views of entities in $S$ in the execution of the real protocol.

    $\Hyb{1}$ (\textit{@User}): In this hybrid, we modify the behavior of simulated honest users $U_x \in U$. Each user $U_x$ now encrypts a randomly selected vector $ V_x$ with the public key $pk_c$ using the Paillier cryptosystem, instead of the original gradient vector $\omega_i$. As only the contents of the ciphertexts are altered, the IND-CPA security property of Paillier~\cite{paillier}, along with the non-collusive guarantees of BS and CSP settings, ensures that this hybrid remains indistinguishable from the previous one.
    
    $\Hyb{2}$ (\textit{@BS}): In this hybrid, we simulate the BS blinding the $\mathcal{W}_i$ with a randomly selected noise $\varepsilon$ rather than $\mu$. The parameters added by uniformly random numbers are also uniformly random. Thus, this hybrid and the previous one are sampled from identical distributions, i.e., uniformly random. In addition, the semantic security property of the Paillier cryptosystem guarantees that this hybrid remains indistinguishable from the previous one.
    
    $\Hyb{3}$ (\textit{@CSP}): In this hybrid approach, we modify the input of Alg.~\ref{alg: PSC} executed by CSP using $\{\chi\}$ instead of $\{{B}\}$. Since only the contents of the ciphertexts are altered, the IND-CPA security of Paillier and the non-collusion assumptions of BS and CSP settings ensure that this hybrid remains indistinguishable from the previous one.

    $\Hyb{4}$ (\textit{@BS}): Similar to \textbf{Hyb{3}}, we change the contents of the output of Alg.~\ref{alg: PSC} and the ciphertext of the obscured gradients. Instead of sending $\varphi$ and $[\mathcal{B}]_{pk_u}$, a random matrix $M_{n\times k}$ and randomly selected vectors $V_i$ are sent to the BS. BS cannot decrypt data that are identical in distribution with the hybrid. Thus, this hybrid is indistinguishable from the previous one.

    The argument proves that there is a simulator \textbf{$SIM$} sampled from the distribution described above so that its output is computationally indistinguishable from the output of \textbf{$REAL$}. Hence, \scheme holds the security property that curious BS and CSP do not learn about the private data of users.
\end{proof}
\begin{proposition}
    \label{the: pi}
        Given a permutation function $\pi$ that maps values from a domain $D$ to the same domain $D$, there is a PPT simulator \textbf{$SIM$} whose output is computationally indistinguishable from a random vector.
    \end{proposition}
    \begin{proof}
        To prove the above theorem, we use a hybrid-dependent argument.
        
        $\Hyb{0}$: The CSP receives the order of the original gradient vectors $v = (v_1, v_2, \ldots, v_n)$.

        $\Hyb{1}$: In this hybrid, the first elements of the vector are permuted using $\pi$, while the remaining elements are left unchanged. The view of CSP is 
        \begin{equation}
            \nonumber
            v' = (\pi(v_1), v_2, \ldots, v_n).
        \end{equation}

        $\Hyb{2}$: In this hybrid, all elements of the vector are permuted using $\pi$. The view of CSP turns to 
        \begin{equation}
            \nonumber
            v'' = (\pi(v_1), \pi(v_2), \ldots, \pi(v_n)).
        \end{equation}
        To prove computational indistinguishability, it is equivalent to prove \textbf{Hyb}$_i$ is indistinguishable from \textbf{Hyb}$_{i+1}$ for $i = 0, 1, \ldots, n-1$. Consider the CSP has a vector 
        \begin{equation}
            \nonumber
            v^i = (\pi(v_1), \pi(v_2), \ldots, \pi(v_i), v_{i+1}, \ldots, v_n)
        \end{equation}
        in \textbf{Hyb}$_i$. Since $\pi$ is a permutation function that provides a random mapping, the output $\pi(v_{i+1})$ is uniformly distributed over $D$. Therefore, $v'$ is computationally infeasible for the CSP to distinguish whether $v_{i+1}$ has been permuted or not without additional information about $\pi$. Along with $\pi$ is a secure permutation function, the distributions of $v^i$ and $v^{i+1}$ are computationally indistinguishable. Thus, $Hyb_i \approx_c Hyb_{i+1}$. By transitivity of computational distinguishability, we have 
        \begin{equation}
            \nonumber
            Hyb_0 \approx_c Hyb_1 \approx_c Hyb_2 \approx_c \ldots \approx_c Hyb_n.
        \end{equation}
        Therefore, $Hyb_0$ (the original vector) is computationally indistinguishable from $Hyb_n$ (the permuted vector). Hence, sending the vector after applying $\pi$ to the server ensures that the vector is computationally indistinguishable from a random vector.
    \end{proof}
\section{Evaluation}
\label{sec:eva}

In this section, we conduct experiments with real-world datasets to evaluate the performance of the \scheme. We utilize Pytorch 1.13.1 to build our machine-learning solution. 
All experiments are run on servers with Intel(R) Xeon(R) Platinum 8362 2.80GHz CPU, 45GB RAM, and NVIDIA GeForce 3090 GPU.

\subsection{Experiment Configuration Details}
\textbf{Datasets.} 
We evaluate \scheme using datasets: MNIST, Fashion-MNIST~\cite{xiao2017fashion}, CIFAR-10~\cite{krizhevsky2009learning}, and CIFAR-100. Both MNIST and  Fashion-MNIST contain 10 categories for classification. CIFAR-100 is extended to 100 classes based on the CIFAR-10; each class has 600 color images. Referring to Hsu \textit{et al.}~\cite{hsu2019measuring}, we use the Dirichlet distribution (note as $Dir(\beta)$) to simulate the non-IID client dataset, where a smaller $\beta$ indicates greater heterogeneity.

\textbf{Model.}
For MNIST and Fashion-MNIST, we use a multilayer perception architecture with two hidden layers, each containing 200 and 100 neurons, respectively. For CIFAR-10 and CIFAR-100, we use a convolutional neural network model consisting of two pairs of a $5\times5$ convolutional layer followed by a $2\times2$ pooling layer, one fully connected layer with 512 units activated by ReLu, and a softmax layer. 

\textbf{Hyper-parameter setting.}
We adopt the notation in~\cite{FedAvg}: $B$ represents the batch size, and $E$ is the number of local epochs. To simulate device heterogeneity among users, we first set the maximum local epoch $E_m = 5$. For each user, we randomly select $n$ numbers within the range $[0.2, 1]$ to simulate $uplink time$, denoted $d$. The local epoch of each user $E_i$ is then calculated as $E_i = d \times E_m$. In our implementation, $B=50$ and $E$ is distributed in $[1, 5]$. The learning rate varied between datasets, $\eta = 0.01$ for MNISTs and $\eta = 0.001$ for CIFARs. We simulate a large volume of participating edge devices in real-world FL with 100 users. 
If not specified, the clustering parameters are $\gamma = 0.5$ and $k = 5$.

\textbf{Baselines.}
We considered the existing techniques FedAvg~\cite{FedAvg}, FedProx~\cite{FedProx}, IFCA~\cite{ghosh2020efficient} and SCAFFOLD~\cite{SCAFFOLD}to evaluate and compare performance.~\cref{sec:rw} discusses the details of the baselines considered.

\begin{figure*}[!ht]
\begin{center}
\centerline{\includegraphics[width=2\columnwidth]{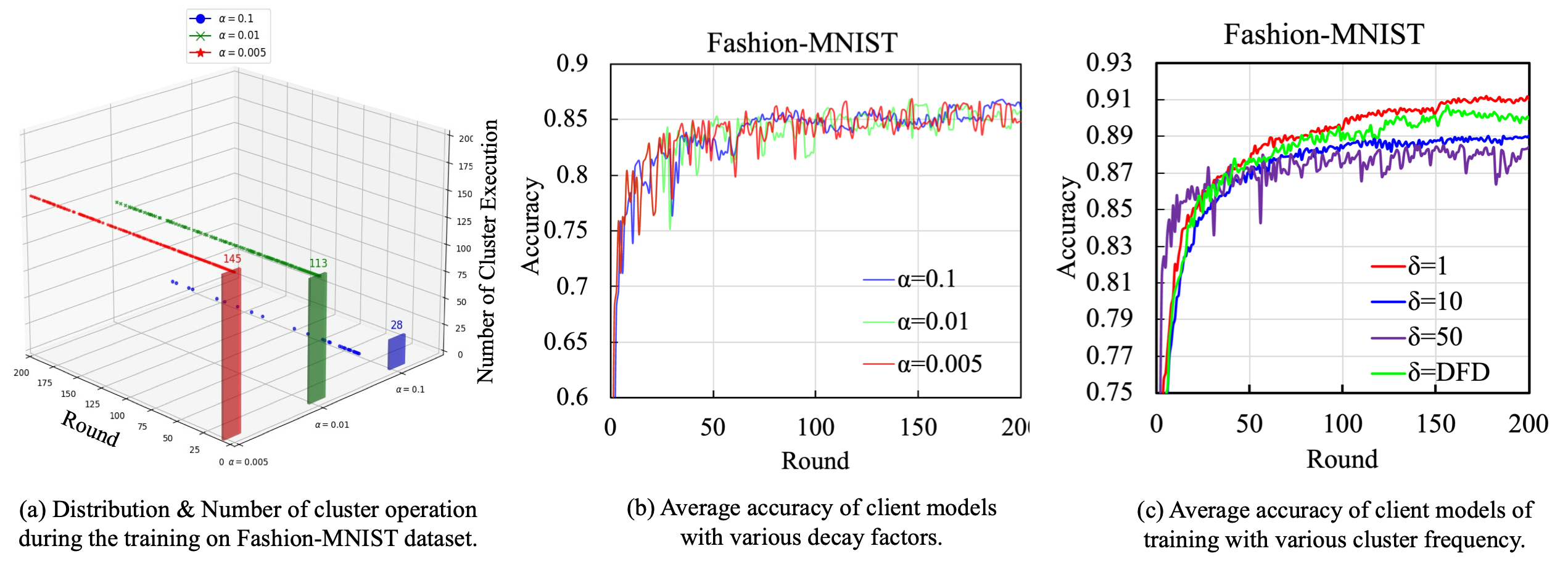}}
\caption{Influence of the dynamic adjustment of cluster frequency. The results are averaged from three independent experiments.}
\label{fig:eod}
\end{center}
\end{figure*}

\begin{figure*}[!ht]
\begin{center}
\centerline{\includegraphics[width=2.1\columnwidth]{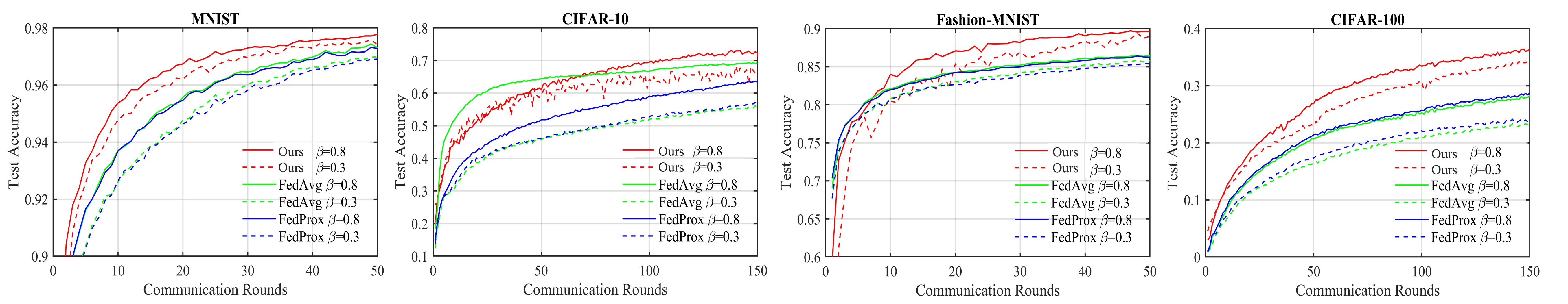}}
\caption{Influence of the heterogeneity on the model optimization in the case of heterogeneous resources.}
\label{fig:hete-acc}
\end{center}
\end{figure*}

\begin{figure}[!ht]
\begin{center}
\centerline{\includegraphics[width=\columnwidth]{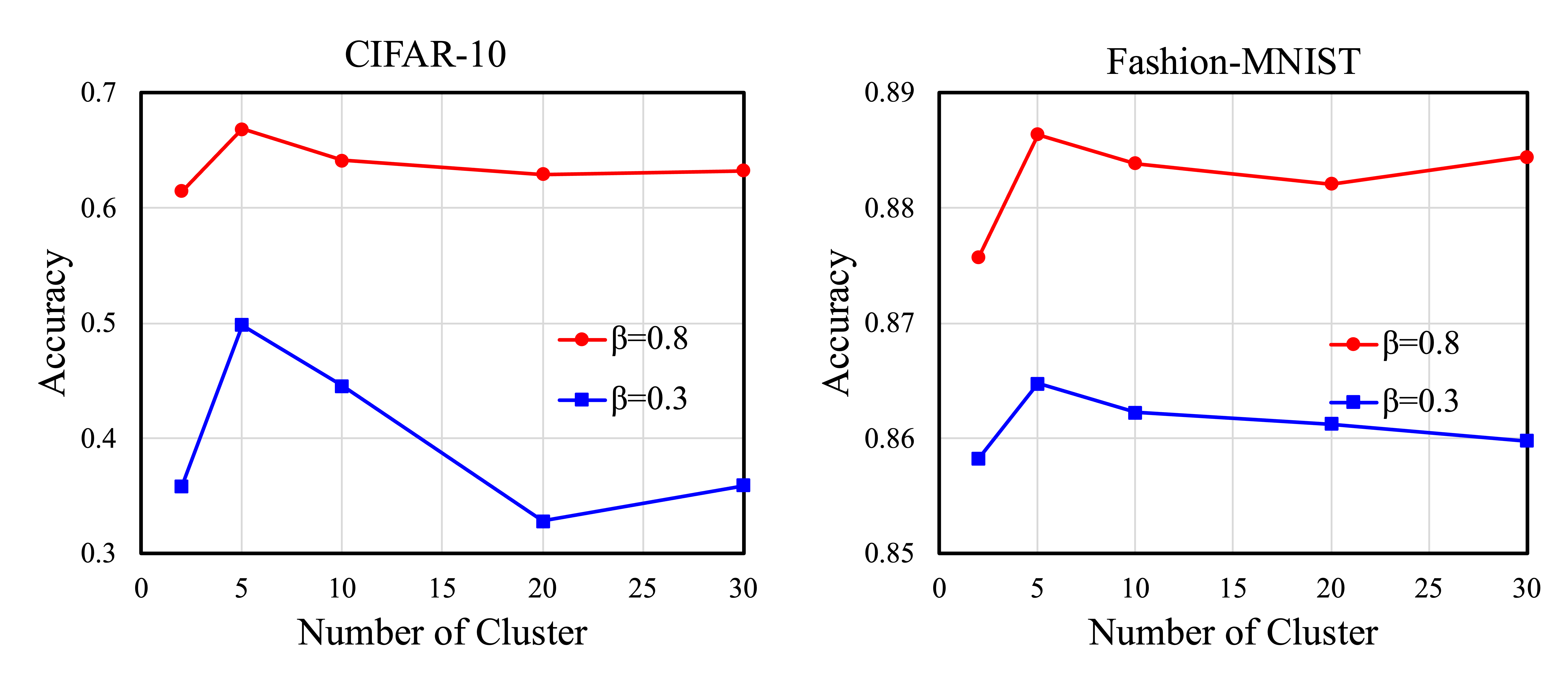}}
\caption{Influence of the number of clustered classes.}
\label{fig:k}
\end{center}
\end{figure}

\begin{table*}[!htb]
\caption{The number of training rounds required to achieve a target test accuracy (\%). The target accuracies are 95\%, 85\%, 65\%, and 33\%, respectively. X denotes the transmission cost if a method cannot achieve the target accuracy. $\mu=0.01$ in FedProx and $k=3$ in ICFA. The results are from three independent experiments.}
\label{tab:comparsion}
\begin{center}
\begin{small}
\begin{sc}
\begin{tabular}{cc|ccccc}
\toprule
  \multirow{2}{*}{Distribution} & \multirow{2}{*}{Dataset} & \multicolumn{5}{c}{Algorithm} \\
        \cmidrule(lr){3-7}
         & & Ours & FedAvg & FedProx & ICFA & SCAFFOLD \\
        \midrule
        \multirow{4}{*}{$Dir(0.8)$} & MNIST & \textbf{9}(95.09\%) & 17(95.03\%) & 18(95.03) & 37(95.21) & 16(\textbf{96.22}) \\
        & Fashion-MNIST & \textbf{12(85.26)} & 28(85.03) & 32(85.18) & 37(85.05) & 29(85.15) \\
        & CIFAR-10 & \textbf{64}(65.41) & 95(65.01) & 96(\textbf{65.42}) & 97(65.13) & 82(66.01) \\
        & CIFAR-100 & \textbf{72}(34.30) & 175(\textbf{34.63}) & 171(34.34) & 115(34.25) & 132(36.25) \\
        \midrule
        \multirow{4}{*}{$Dir(0.3)$} & MNIST & \textbf{13(95.46)} & 23(95.18) & 23(95.04) & 45(95.03) & 20(95.15) \\
        & Fashion-MNIST & \textbf{18(85.17)} & 40(85.07) & 45(85.09) & 62(85.07) & 38(85.13)\\
        & CIFAR-10 & \textbf{93(65.61)} & 157(65.59) & 159(65.01) & 233(65.04) & 116(65.06) \\
        & CIFAR-100 & \textbf{99}(33.16) & 201(33.56) & 200(33.23) & X(29.84)& 115(\textbf{33.70}) \\
\bottomrule
\end{tabular}
\end{sc}
\end{small}
\end{center}
\end{table*}

\begin{table}[!htb]
\caption{Impact of the degree of resource heterogeneity on the model accuracy (\%) $\pm$ std. The results are averaged from three independent experiments.}
\label{tab:hete}
\begin{center}
\begin{small}
\begin{sc}
\resizebox{0.5\textwidth}{!}{
\begin{tabular}{lcccr}
\toprule
Dataset & Algorithm &$Dir(0.8)$ & $Dir(0.3)$  & $Dir(0.1)$ \\
\midrule
\multirow{2}{*}{MNIST}  & Ours& \textbf{98.78$\pm$0.25}  & 98.38$\pm$0.45  & 97.33$\pm$0.30 \\
                            & FedAvg & 97.94$\pm$0.30   & 97.69$\pm$0.35  & 96.44$\pm$0.25 \\
                            \midrule
    \multirow{2}{*}{FMNIST} &  Ours &\textbf{90.89$\pm$0.20}  &90.12$\pm$0.35   &89.37$\pm$0.40 \\
                            & FedAvg &87.69$\pm$0.35   &86.88$\pm$0.25   &82.91$\pm$0.30   \\
                            \midrule
    \multirow{2}{*}{CIFAR-10}  & Ours & \textbf{74.86$\pm$0.40}   & 68.01$\pm$0.25   & 64.43$\pm$0.50  \\
                             & FedAvg& 69.33$\pm$0.30   & 65.59$\pm$0.25  & 45.59$\pm$0.25   \\
                            \midrule
    \multirow{2}{*}{CIFAR-100} & Ours &\textbf{40.07$\pm$0.40}   &36.05$\pm$0.40   &32.25$\pm$0.35   \\
                            & FedAvg &37.06$\pm$0.25   &33.56$\pm$0.30   &23.56$\pm$0.30  \\
\bottomrule
\end{tabular}
}
\end{sc}
\end{small}
\end{center}
\end{table}

\subsection{Ablation Studies}
\subsubsection{Effects of DFD Strategy}
To investigate the decay rate of clustered frequencies on model accuracy, we conduct experiments on the Fashion-MNIST datasets in the Dirichlet(0.3) setting. As shown in~\cref{fig:eod}(a), with decay factors $\alpha$ of 0.1, 0.01, and 0.005, DFD reduces clustering costs by $1.28\times$ to $7.14\times$ compared to $\alpha=1$ (Clustering occurs in each round). Moreover, $67.9\%$ of the cluster operations arise early in the training with $\alpha=0.1$. Furthermore, as depicted in~\cref{fig:eod}(b), the model accuracy oscillations increase as the decay factor decreases, suggesting that the lower clustering frequency does not hinder the model convergence in heterogeneous environments. We also compare the accuracy of the model under different cluster frequency settings. In~\cref{fig:eod}(c), the DFD method results in $1.03\%$ lower accuracy than $\delta=1$, while cluster costs drop by $86\%$.~\cref{fig:eod} supports the idea that excessive clustering frequency can disrupt convergence and lead to unnecessary computations.

\subsubsection{Effects of Group Strategy}
To evaluate the effect of the number of clusters on the model accuracy results, we conduct experiments on the Fashion-MNIST and CIFAR-10 datasets. We set hyperparameters as $\alpha=0.1, \beta=0.3$ for DFD strategy and non-IID settings, respectively. As demonstrated in~\cref{fig:k}, when the number of groups is 2, 5, 10, 20, and 30, respectively, the accuracy of the model initially increases and then decreases. This indicates that if the number of clustered classes is insufficient, it becomes challenging to group models trained with homogeneous datasets, resulting in a deterioration of model accuracy. Conversely, if the number of clusters is excessively large, the heterogeneity among clusters increases, and a cluster model provides less data information to the local model.
As the scheme ensures that the local model can learn from multiple cluster models, the heterogeneity of cluster models also impacts the final model's accuracy.

\subsection{Performance Comparison}
We evaluated communication efficiency in two ways: resource heterogeneity and task difficulty.~\cref{tab:comparsion} shows that \scheme achieves optimal communication efficiency in all settings. The model trained in \scheme achieves predetermined accuracy with fewer communication rounds.

\textbf{Impact of the non-IID data.} The data distributions between users exhibit more non-IID as we transition from $Dir(0.8)$ to $Dir(0.3)$ settings, leading to a decrease in model accuracy and a slower convergence rate. From the setting of $Dir(0.8)$ to $Dir(0.3)$, the average number of communication rounds with the server required by FedAvg to achieve the target accuracy is $2.319\times$ and $2.008\times$ than that of \scheme, respectively. For ICFA, a clustering FL method, the required average number of communication rounds is $2.474\times$ that of \scheme under slight heterogeneity, and the constant clustering strategy struggles to tackle the diversity of data distribution across nodes.

\textbf{Impact of task difficulty.} We first concentrate on the scenario where the non-IID setting is $Dir(0.8)$. The variance of the final communication rounds for \scheme in four tasks is 20.63\%, 22.36\%, 66.69\%, and 38.74\% of the corresponding values for the other four FL methods. This indicates that \scheme is more robust to task difficulty via adaptive clustering strategies to reduce the adverse effects of resource heterogeneity. Compared to FedAvg, the final training accuracy of \scheme is 0.79\%, 3.34\%, 2.42\%, and 2.59\% higher for the four datasets when the setting is $Dir(0.8)$.

\textbf{Impact of heterogeneity.}~\cref{tab:hete} illustrates the impact of resource heterogeneity on the model performance of \scheme. When $\beta$ is set to 0.8, 0.3, and 0.1, the accuracy of the FedAvg model shows a significant declining trend across four datasets, with this trend being more pronounced in hard tasks (\textit{i.e.}, CIFARs), as illustrated in~\cref{fig:hete-acc}. In contrast, the accuracy of the \scheme model is less affected by heterogeneity across all datasets. This can be attributed to the fact that as resource heterogeneity increases, the soft clustering strategy leverages more data information from different devices, ultimately enhancing the accuracy of the final model.

\section{conclusion}
\label{sec:clu}
We propose \scheme, a privacy-preserving heterogeneous federated learning framework with anonymous adaptive clustering to protect the privacy of client data and identity under the heterogeneous environment. \scheme mitigates the negative impact of the heterogeneous problem and achieves the target accuracy with fewer communication rounds. Experimental results in various heterogeneous environments demonstrate that \scheme exhibits consistently high efficiency and effectiveness. 

\bibliographystyle{IEEEtran}
\bibliography{ref}


%





\ifCLASSOPTIONcaptionsoff
  \newpage
\fi

\end{document}